# Optimal Node Selection using Estimated Data Accuracy Model in Wireless Sensor Networks

Jyotirmoy karjee, H.S Jamadagni
Department of Electronic Systems Engineering
Indian Institute of Science
Bangalore, India

***Abstract*—One of the major task of wireless sensor network is to sense accurate data from the physical environment. Hence in this paper, we develop an estimated data accuracy model for randomly deployed sensor nodes which can sense more accurate data from the physical environment. We compare our results with other information accuracy models and shows that our estimated data accuracy model performs better than the other models. Moreover we simulate our estimated data accuracy model under such situation when some of the sensor nodes become malicious due to extreme physical environment. Finally using our estimated data accuracy model we construct a probabilistic approach for selecting an optimal set of sensor nodes from the randomly deployed maximal set of sensor nodes in the network.**

## I. INTRODUCTION

Recent progress in wireless technology has made a drastic improvement over wireless sensor networks. In wireless sensor networks, nodes are deployed in the sensing region to sense the physical phenomenon from the environment and transmit the data to the sink node. The sensor nodes sense the physical phenomenon of data for the event like seismic event, fire, temperature, humidity etc. from the physical environment [1]. Data collected by the sensor nodes are spatially correlated [2] among them in the sensing region. These spatially correlated data sensed by the sensor nodes are directly transmitted to the sink node which estimates [3,4,5,6] the data (information) accuracy. In the literature [7,8,9,10],authors discuss the data accuracy under distributed conditions. In this paper, we take the same scenario discussed in [10] where we deploy sensor nodes randomly in the sensor region. When a source event occurs, the maximal set of sensor nodes wake up to sense the source event, then transmits the sensed data to the sink node. The main motivation of these paper is to develop an estimated data accuracy model to sense more accurate data from the physical environment, we compare our results with the other information accuracy (distortion) models [4,5,6]. Finally a probabilistic model is proposed to select an optimal set of sensor nodes to sense the data from the randomly deployed maximal set of sensor nodes in the network using data accuracy function [7]. In literature [17] maximizing the network life time subjected to event constraint and in literature [18] total information gathered subjected to energy constraints are discussed without verifying the information accuracy. Hence gathering or aggregating information without verifying the accuracy level cause problem if some of the sensor nodes get malicious [11]. If the sensor nodes get malicious, it can read inaccurate data and transmits the inaccurate data to the sink node. If the inaccurate data gets aggregated with the other correct data sensed by the sensor nodes, it causes incorrect data aggregation at the sink node. Hence sink node estimate the incorrect data reading for the network. Therefore it is essential to verify the data accuracy before data aggregation discussed in [7,8,9,10].However to the best understanding of authors , this is the first time we perform results for data accuracy when some of the sensor nodes behaves live malicious nodes due to extreme physical environment such as heavy rain fall , heavy snow fall in the hilly region etc.

The rest of the paper is given as follows. In the section II, we demonstrate a system model where we explained the node deployment strategy and constructed the theoretical model for estimated data accuracy under spatially correlated data constraints. In section III we perform the simulation for estimated data accuracy model under various topological scenarios. In section IV, we perform a probabilistic model to find the optimal set of sensor nodes in the network using data accuracy function and finally we conclude our work in the section V.

## II. SYSTEM MODEL

In this section, we discuss the node deployment strategy in the wireless sensor network and illustrate how a set of sensor nodes perform data accuracy at the sink node. Finally we develop a model for estimated data accuracy for the set of sensor nodes measured at the sink node.

*Basic notations used in estimated data accuracy model:*

$U$ = Set of sensor nodes deployed
$\|V\| = v$ = Total number of wake up nodes
$\|W\| = w$ = Number of optimal nodes selected
$S$ = Source event
$\hat{S}$ = Estimation of source event
$X_i$ = Observation made by $i$ sensor node

*A. Deployment of Nodes in the Sensing Region*

Assuming $U$ set of sensor nodes deployed randomly over a region $Q$ such that $Q \subseteq R^2$ where $\|U\| = u$ are the total number of sensor nodes. Suppose a source event $S$ has occurred in the sensing region $Q$. Source event [10] is an event that originates at a point and radiates outwards e.g fire. When $S$ occurred in $Q$, a set of sensor nodes $V$ wake up to sense the physical phenomenon of $S$. We define $V$ as maximal set of sensor nodes which forms fully connected network among them. A fully connected network [10] is a network topology in which there exists a direct one hop link between all pair of sensor nodes. We define $\|V\| = v$ be the number of sensor nodes where $v(v-1)/2$ is the direct one hop link to form a fully connected network with $v$ sensor nodes. Finally $V$ set of sensor nodes wake up to sense the physical phenomenon of source event and transmit the observed data to the sink node explained below.

*B. Model for Estimated Data Accuracy in the Network*

Since $V$ set of sensor nodes wake up in the network to sense the physical phenomenon of source event $S$, we construct a mathematical model to estimate the observed data at the sink node. Sink node is responsible for collecting the observation made by $v$ sensor nodes to estimate $\hat{S}$ from $S$. Hence the error signals [ 13,14 ] can be defined as

$$\tilde{S} \triangleq (S - \hat{S}) \tag{1}$$

Our intuition is to determine $\hat{S}$ by minimizing the mean square error from the expectation of $\tilde{S}^2$.

$$\min_{\hat{S}} E(\tilde{S})^2 \tag{2}$$

The observation made by each sensor node $i$ is given as

$$X_i = S_i + N_i \quad \text{where } i \in V \tag{3}$$

We assume uncoded transmission for the observed data sensed by sensor nodes in the network. Each sensor node $i$ transmits a scaled version $Y_i$ [4] of the observed data $X_i$ to the sink node with power constraint $P$. Hence transmitted signal is given as

$$Y_i = \sqrt{\frac{P}{\sigma_{S_i}^2 + \sigma_{N_i}^2}} X_i$$

Expressing $\alpha_i = \sqrt{\frac{P}{\sigma_{S_i}^2 + \sigma_{N_i}^2}}$ where $i \in V$

Hence $Y_i = \alpha_i X_i$

The encoded signal $Y_i$ transmitted by each sensor node $i$ through additive white Gaussian noise (AWGN) channel [6,12] is sent to the sink node. Sink node store the received signal in $Y$ matrix for all sensor nodes as

$$Y = \alpha X \tag{4}$$

$$\text{for } Y = \begin{pmatrix} Y_1 \\ Y_2 \\ . \\ Y_v \end{pmatrix}, \alpha = \begin{pmatrix} \alpha_1 & 0 & . & 0 \\ 0 & \alpha_2 & . & . \\ . & . & . & . \\ 0 & . & . & \alpha_v \end{pmatrix} \text{ and } X = \begin{pmatrix} X_1 \\ X_2 \\ . \\ X_v \end{pmatrix}$$

where $i = 1, 2.........v$

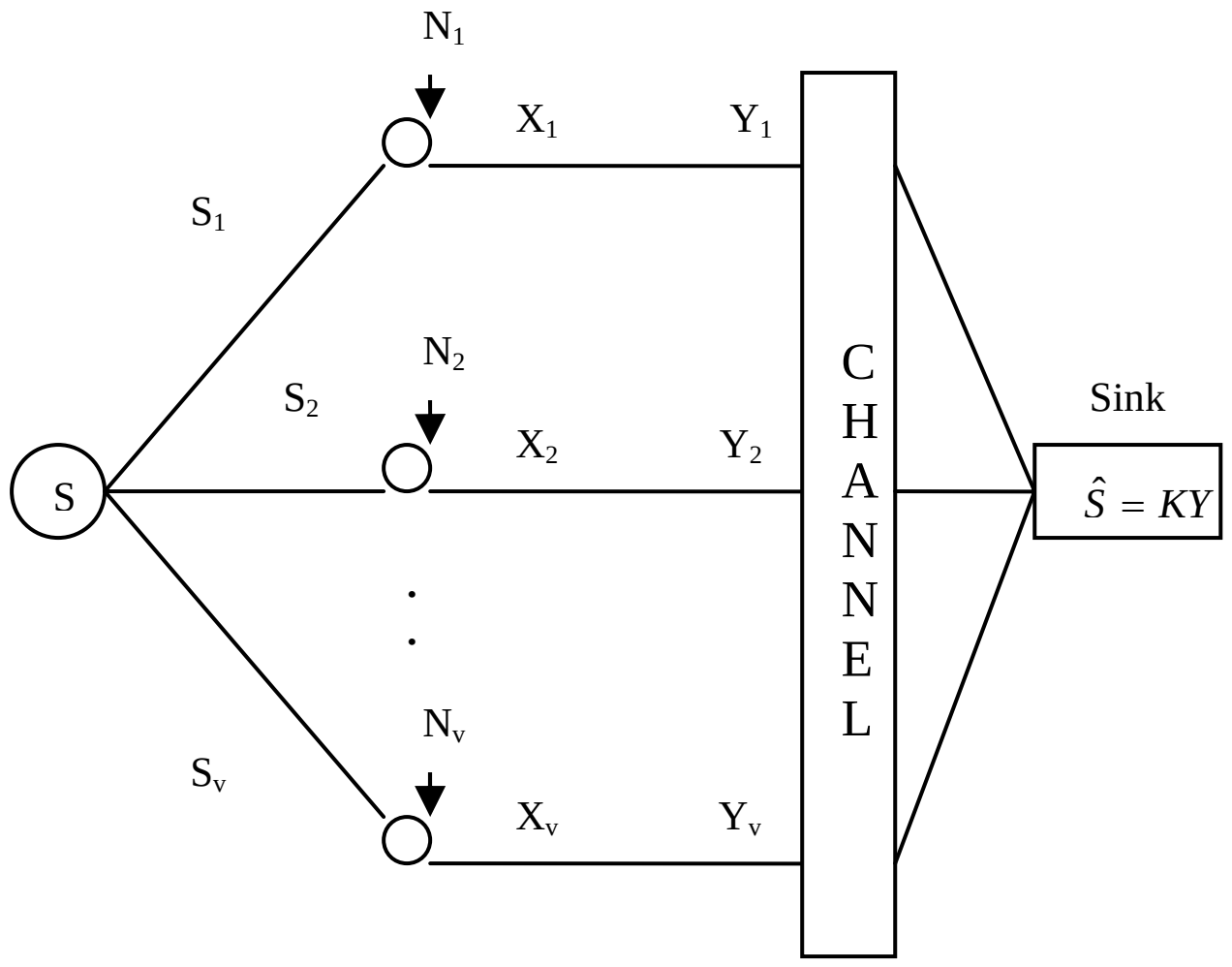

Figure 1: Architecture for Data Estimation Model in WSNs

Sink node decodes the signal to retrieve the estimate $\hat{S}$ of $S$. We define $\hat{S}$ as a random variable as a function of $Y$ to recover the estimate $\hat{S}$ of all the observations done by sensor nodes at the sink node.

$$\hat{S} = h(Y) \tag{5}$$

Thus the mean square error becomes

$$\min_{h(Y)} E(\tilde{S})^2 \tag{6}$$

We choose $h(Y)$ for the subclass of affine functions [14 ] of Y

$$h(Y) = (KY + b)$$

Where $K$ is matrix and $b$ is a vector. Hence

$$\hat{S} = (KY + b) \tag{7}$$

The affine estimator of $S$ is unbiased, hence we have

$$E(\hat{S}) = 0$$

$$\text{For} \quad E(\hat{S}) = KE(Y) + b = b$$

For a linear estimator, we have $b = 0$ to get

$$\hat{S} = KY \tag{8}$$

Hence we find the optimal value of $K$ at the sink as shown in Fig.1 for $\hat{S}$ such that

$$\min_K E(S-KY)^2 \quad (9)$$

*Finding the optimal value: We* calculate the optimal value of $K$ for the estimate $\hat{S}$ using orthogonality principle. $Y$ is orthogonal to the error signal $(\tilde{S})$ .i.e $Y \perp \tilde{S} = 0$. To get the optimal value of $K$ for the estimator at the sink node ,we define a linear model for (4) as

$$Y = \alpha(ZS+N) \quad (10)$$

$$Y = \begin{pmatrix} Y_1 \\ Y_2 \\ . \\ Y_v \end{pmatrix} = \begin{pmatrix} \alpha_1 & 0 & . & 0 \\ 0 & \alpha_2 & . & . \\ . & . & . & . \\ 0 & . & . & \alpha_v \end{pmatrix} \left[ \begin{pmatrix} Z_1 \\ Z_2 \\ . \\ Z_v \end{pmatrix} S + \begin{pmatrix} N_1 \\ N_2 \\ . \\ N_v \end{pmatrix} \right]$$

For zero mean random vector $\{S,Y\}$ for some matrix $Z$ . $N$ is a zero mean random noise vector with known covariance matrix $E(NN^T) \approx \sigma_N^2 I$ . The covariance matrix of $S$ is also known $E(SS^T) \approx \sigma_S^2 I$ and $\{S,N\}$ are uncorrelated.

The linear least mean square estimator works according to orthogonality principal as follows

$$E\left(Y^T\tilde{S}\right) = 0$$

$$E\left(Y^T\left(S-KY\right)\right) = 0$$

Hence the optimal value of $K$ is given as

$$K = \frac{E(Y^TS)}{E\left(Y^TY\right)} \quad (11)$$

Substituting (10) in (11) we get the expression as

$$K = \frac{\sigma_S^2 Z^T \alpha^{-1}}{\left(\sigma_S^2 Z^T Z + \sigma_N^2\right)} \quad (12)$$

Using (12) in (8), we get the linear least mean square estimator of $S$ given $Y$ is

$$\hat{S} = \frac{\sigma_S^2 Z^T \alpha^{-1}}{\left(\sigma_S^2 Z^T Z + \sigma_N^2\right)} Y \quad (13)$$

Put the value of (4) in (13), we get

$$\hat{S} = \frac{Z^T}{\left(Z^T Z + \sigma_N^2/\sigma_S^2\right)} X$$

Hence linear least mean square estimator of $S$ given $X$ for $V$ sensor nodes in the network is given as

$$\hat{S}(V) = \frac{1}{\left(V + \sigma_N^2/\sigma_S^2\right)} \sum_{i=1}^{V} X_i$$

$$\hat{S}(V) = \frac{1}{\beta} \sum_{i=1}^{V} X_i \quad (14)$$

Denoting $\beta = \left(V + \dfrac{\sigma_N^2}{\sigma_S^2}\right)$

We define mean square error between $S$ and $\hat{S}(V)$ to find the estimated data accuracy [9] for $V$ sensor nodes in the network as

$$D(V) = E[S - \hat{S}(V))]^2$$

$$D(V) = E[S^2] - 2E[S\hat{S}(V)] + E[\hat{S}(V)^2] \quad (15)$$

The normalized [6,10] data accuracy $D_A(V)$ for $V$ sensor nodes in the network is given as

$$D_A(V) = 1 - \frac{D(V)}{E[S^2]}$$

$$D_A(V) = \frac{1}{E[S^2]}[2E[S\hat{S}(V)] - E[\hat{S}(V)^2]] \quad (16)$$

The normalized data accuracy $D_A(V)$ for $V$ sensor nodes in the sensor region can be implemented in spatial correlation model explained in the next part.

*C. Estimated Data Accuracy Model for Spatially Correlated Data*

We demonstrate a mathematical model for the normalized data accuracy model for spatial correlated data among $V$ sensor nodes in the sensing region. $V$ sensor nodes sense the observed data for the source event $S$ in the sensing region. We model a spatially correlated physical phenomenon of sensed data for $V$ sensor nodes as a joint Gaussian random variable (JGRV's) [4,5] as follows:

Step 1: $E[S]=0$ , $E[S_i]=0$ , $E[N_i]=0$

Step 2: $Var[S] = E[SS^T] = \sigma_S^2$ , $Var[S_i] = E[S_iS_i^T] = \sigma_{S_i}^2$

$Var[N_i] = E[N_iN_i^T] = \sigma_{N_i}^2$

Step 3: $Cov[S,S_i] = \sigma_S^2 Corr[S,S_i]$ , $Cov[S_i,S_j] = \sigma_S^2 Corr[S_i,S_j]$

Step 4: $E[S,S_i] = \sigma_S^2 Corr[S,S_i] = \sigma_S^2 \rho(S,S_i) = \sigma_S^2 K_v(d_{S,S_i})$

$E[S_i,S_j] = \sigma_S^2 Corr[S_i,S_j] = \sigma_S^2 \rho(S_i,S_j) = \sigma_S^2 K_v(d_{S_i,S_j})$

We illustrate the covariance model [15] for Step 3-4 for spatially correlated data among sensor nodes in the network. Using covariance model we have $K_V(d_{i,j})$ where $d_{i,j} = \| S_i - S_j \|$ represents the Euclidian distance between node $i$ and $j$ .The covariance function is non-negative and decrease monotonically with the Euclidian distance $d_{i,j} = \| S_i - S_j \|$ with limiting values of $1$ at $d=0$ and of $0$ at $d=\infty$ . We take the power exponential model [16] i.e $K_V^{P.E}(d_{i,j}) = e^{-(d_{i,j}/\theta)}$ for $\theta > 0$ where $\theta$ is called as 'Range parameter'. 'Range parameter' controls the relation between the distance among sensor nodes $(i, j)$ and the correlation coefficient $\rho(i, j)$. Thus from the correlation model, we get $\rho_{S_i,S} = e^{-(d_{S,i}/\theta)}$ and $\rho_{S_i,S_j} = e^{-(d_{i,j}/\theta)}$.

Using (3) and (14) in (16), we get the estimated normalized data accuracy for $V$ sensor nodes in the network as

$$D_A(V) = \frac{1}{\beta}\left(2\sum_{i=1}^{V} e^{-(d_{S,i}/\theta)}\right) - \frac{1}{\beta^2}\left(\sum_{i=1}^{V}\sum_{j=1}^{V} e^{-(d_{i,j}/\theta)}\right) - \left(\frac{\left(\sum_{i=1}^{V}\sigma_{N_i}^2\right)}{\beta^2\sigma_S^2}\right) \quad (17)$$

## III. Simulation Results

To perform the simulations, we deploy $U$ set of sensor nodes in the sensor region $Q$. When a source event is detected, $V$ set of sensor nodes wake up with fully connected network from $U$ set of sensor nodes in the sensing region. In the first simulation, we take four sensor nodes $(v = 4)$ with fully connected network which can sense a source event. We vary the distance from the sensor nodes to the source event with fixed $(v = 4)$ sensor nodes i.e $d_{s,i}$ where $i = 1, 2, 3, 4$ are equidistance in the sensing field as shown in Fig. 2. For this we put $v = 4$ sensor nodes in a deployed circle with a source event occurred at the centre of the deployed circle. As we keep on increasing the radius of the deployed circle for $d_{s,i}$ with same proportion, the estimated data accuracy decreases. This means as the distance from the source event $S$ to the $v$ sensor nodes increases, the estimated data accuracy decreases as shown in Fig. 3. We take $\theta = 70$ for our statistical data to calculate the estimated normalized data accuracy and compare the results from the literature [4], [5] and [6]. We conclude that as the radius of the deployed circle increases with same proportion our estimated data accuracy model always perform better compare to other models [4,5,6]. The accuracy models, [5] and [6] shows the same results for the normalized data accuracy with increasing deployed circle as shown in Fig.3.

In the second simulation, a sensing region of $2m \times 2m$ grid based sensor topology is taken with a sink node and a fixed source event occurred in the center of the sensor network as shown in Fig. 4. We deployed thirty four nodes and a sink node in a grid based sensor topology as done in literature [5,8]. When we simulate for this topology, we get the results for estimated data accuracy as shown in Fig. 5 which can sense more accurate data compare to other information (distortion) accuracy model [4,5,6]. Thus our propose estimated data accuracy model can sense more accurate data than the other work done [4,5,6] ,as we keep increasing the number of sensor nodes for $\theta = 70$. Moreover the data accuracy remains approximately constant after a certain saturation level still we continuously increase the number of sensor nodes in the sensing field. In Fig. 5, it is clear that seven to ten sensor nodes are sufficient for achieving the same estimated data accuracy level instead of deploying thirty four sensor nodes in the field. Hence it is unnecessary to choose all the sensor nodes to perform data accuracy at the sink node. Therefore an optimal $W$ set of sensor nodes can be selected for sensing the source event and performs the data accuracy at the sink node.

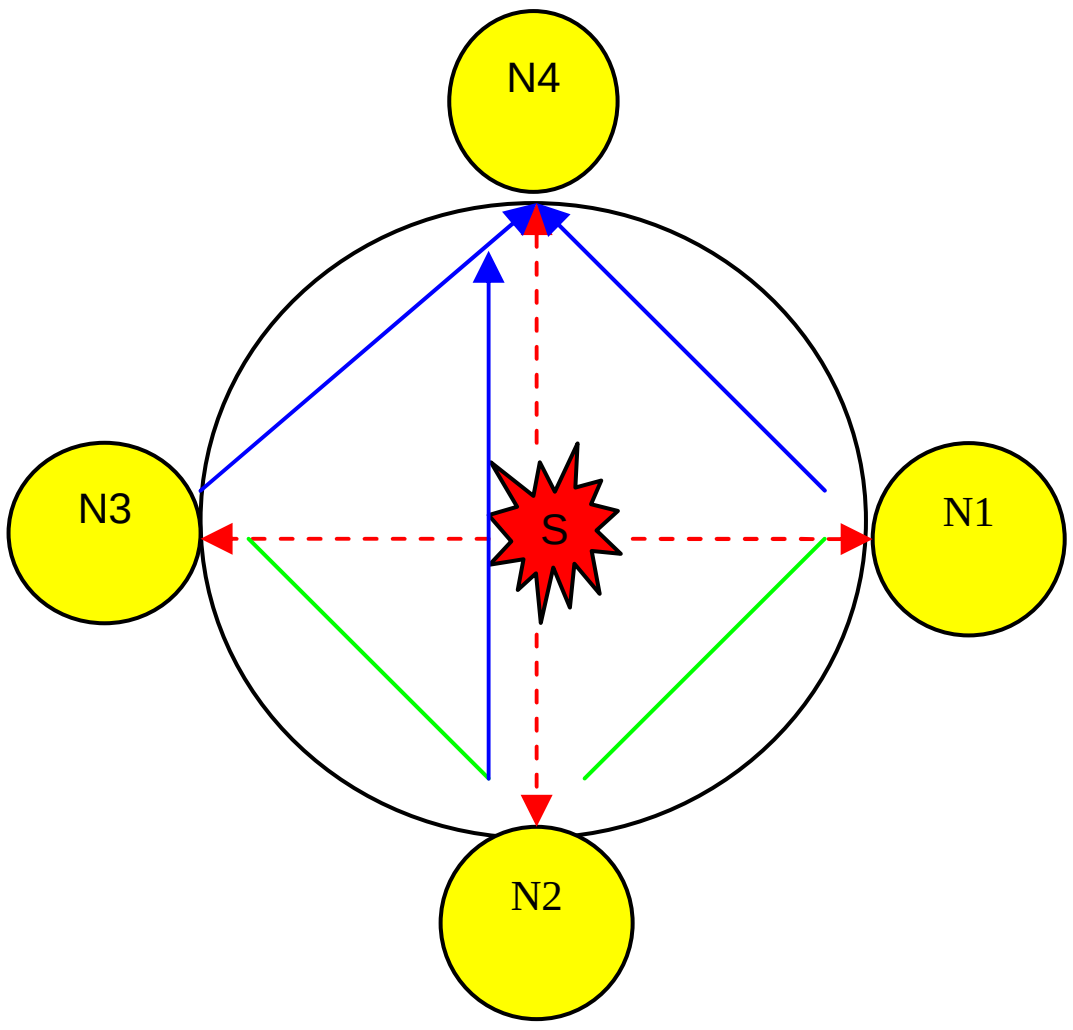

Figure 2: Circular topology for deployed sensor nodes

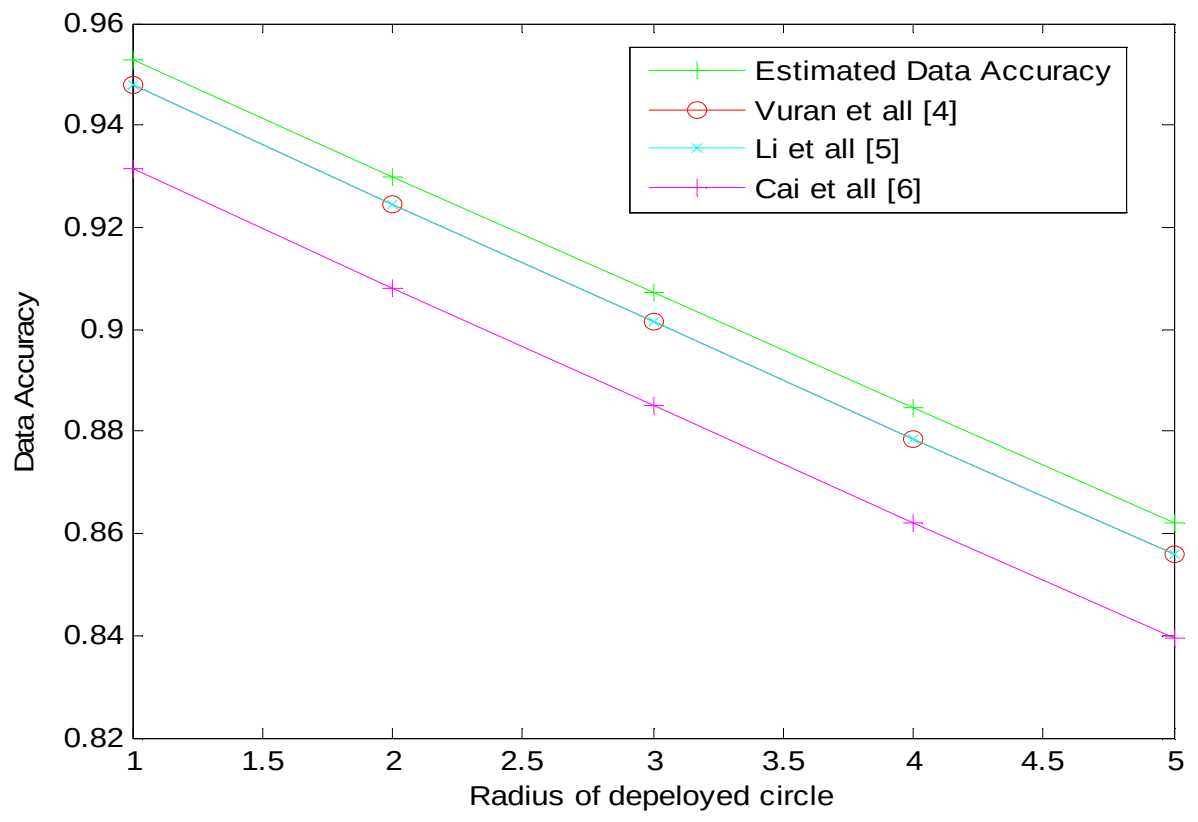

Figure 3: Radius of circular topology versus data accuracy

In the third simulation, we take the same scenario as taken in the previous simulation setup. We assume that some of the sensor nodes get malicious due to extreme physical environment e.g. heavy rain fall. In such tropical situation, these sensor nodes read inaccurate data in the sensing region. This means noise variances of the malicious nodes are much higher compare to the noise variance of the normal nodes. We define normal node as non malicious node or good node in the wireless sensor network. We perform simulation results for estimated data accuracy model and compare with the other information accuracy models [4,5,6] under malicious nodes condition in the network. For the simplicity of our simulation, we initially deployed ten sensor nodes and assume out of ten sensor nodes ,six sensor nodes are malicious as shown in Fig. 6. We keep on adding the number of sensor nodes to thirty four sensor nodes deployed randomly in the network. We estimate data accuracy for thirty four sensor nodes at the sink node and compare with the other information accuracy models under malicious nodes condition in the network. Finally we conclude that our estimated data accuracy model perform much better than other models still we introduce some malicious nodes in the network.

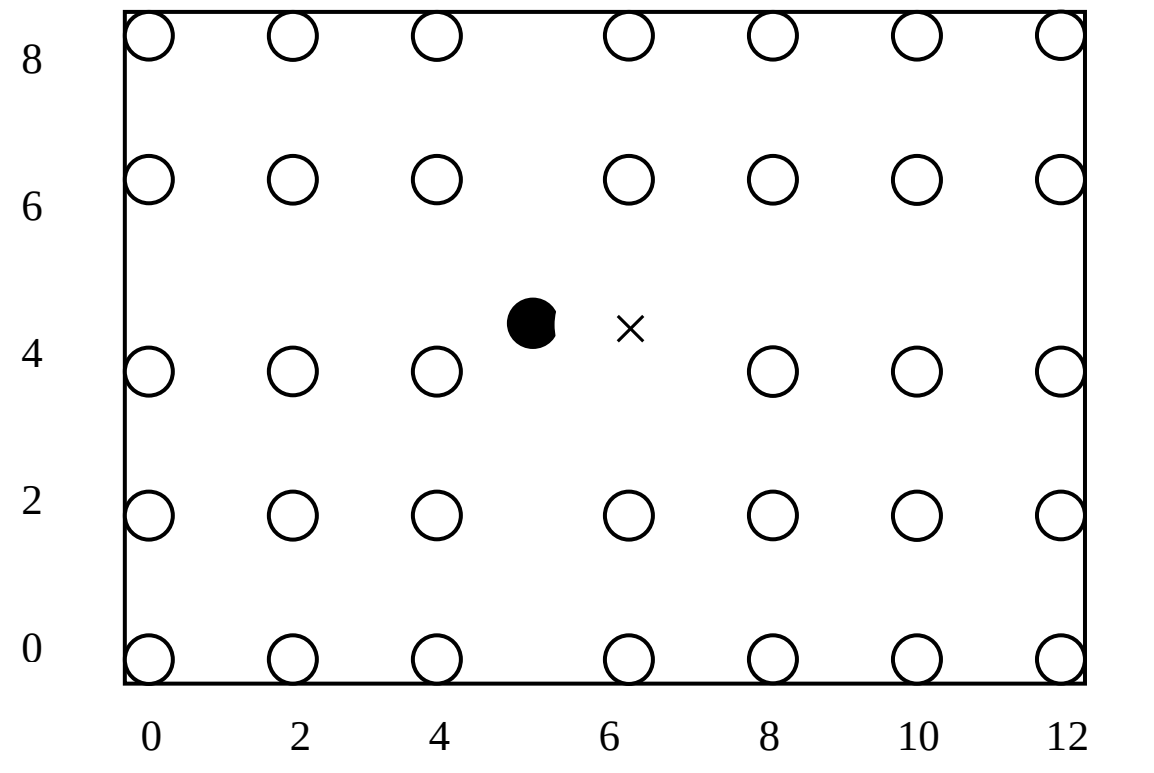

Figure 4: Wireless sensor network topology: ○ means sensor node, ● means sink node, × means source event.

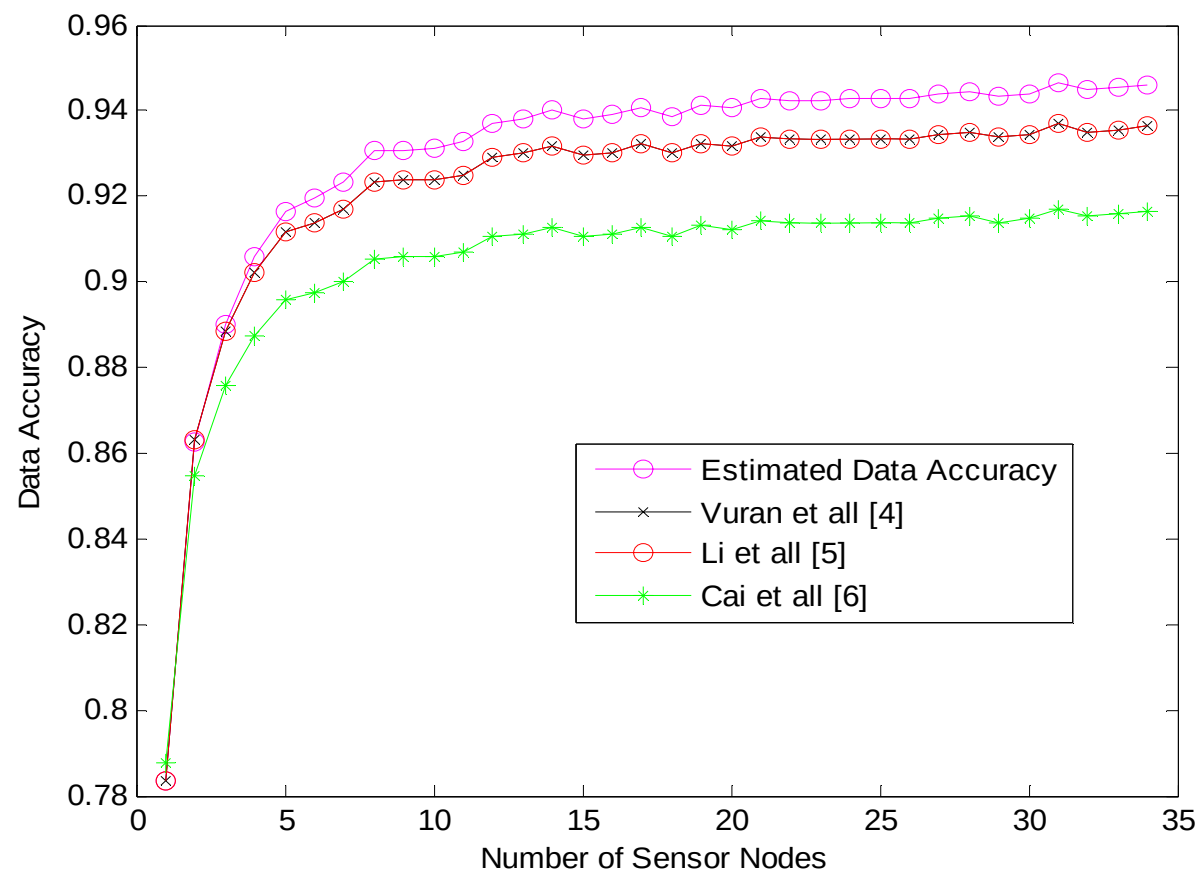

Figure 5: Number of sensor nodes versus data accuracy

In Fig. 7, we simulate for our estimated data accuracy model with respect to normal nodes and for introducing some malicious nodes in the sensing region. In this simulation setup, we compare two deployment strategies. In the first strategy, initially we deploy ten sensor nodes and keep adding sensor nodes to thirty four sensor nodes. These nodes are normal nodes. In another node deployment strategy, initially we deployed ten sensor nodes in similar way but out of ten nodes, six sensor nodes are malicious. We keep going on adding sensor nodes till we get thirty-four sensor nodes in the network. We compare these two deployment strategies for sensor nodes and conclude that sink node estimates more accurate data when there are normal nodes in the network. But if there are some malicious nodes in the network, the sink node estimates inaccurate data and performs poor data gathering for the deployed sensor nodes. Another conclusion we can draw from Fig. 7 is that the effect of noise variances of malicious nodes decreases as we keep increasing the number of sensor nodes in the network.

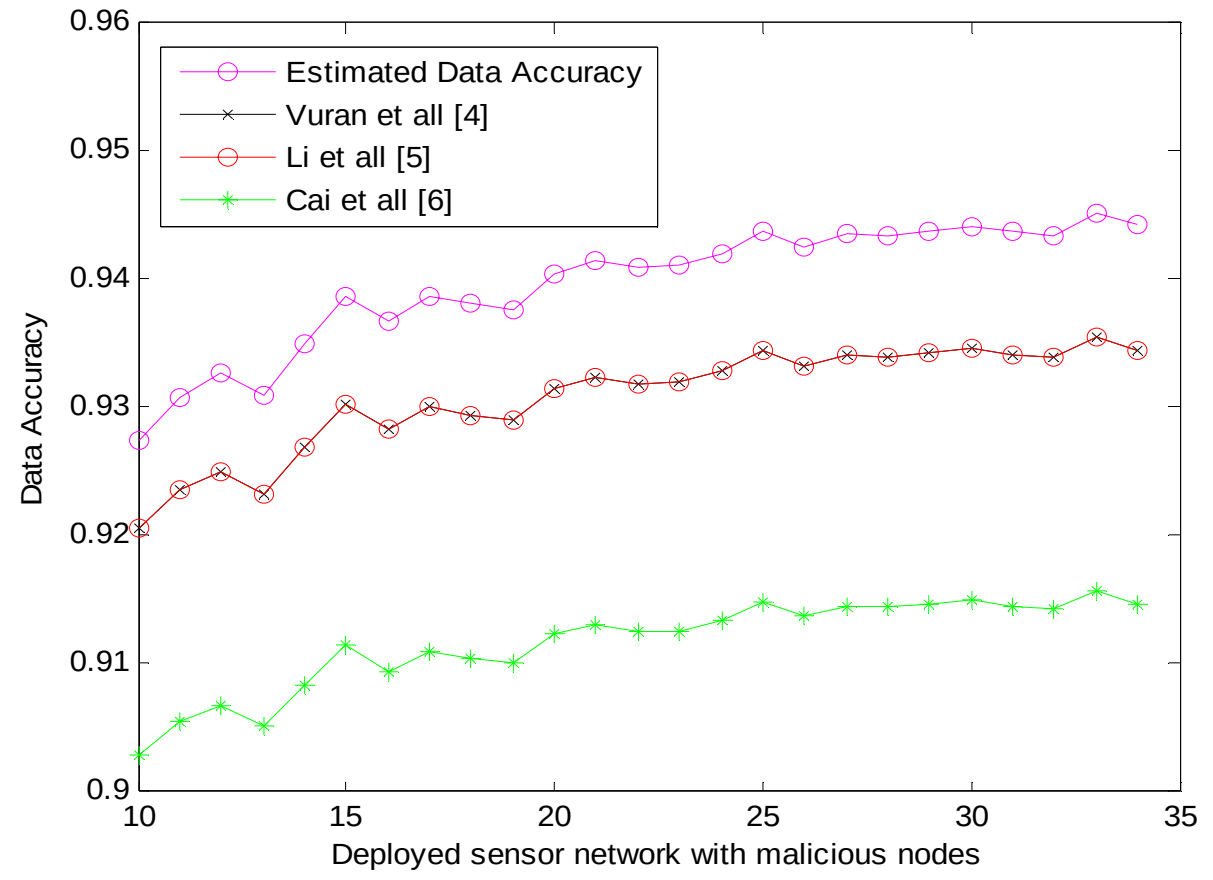

Figure 6: Data accuracy for malicious nodes in the network

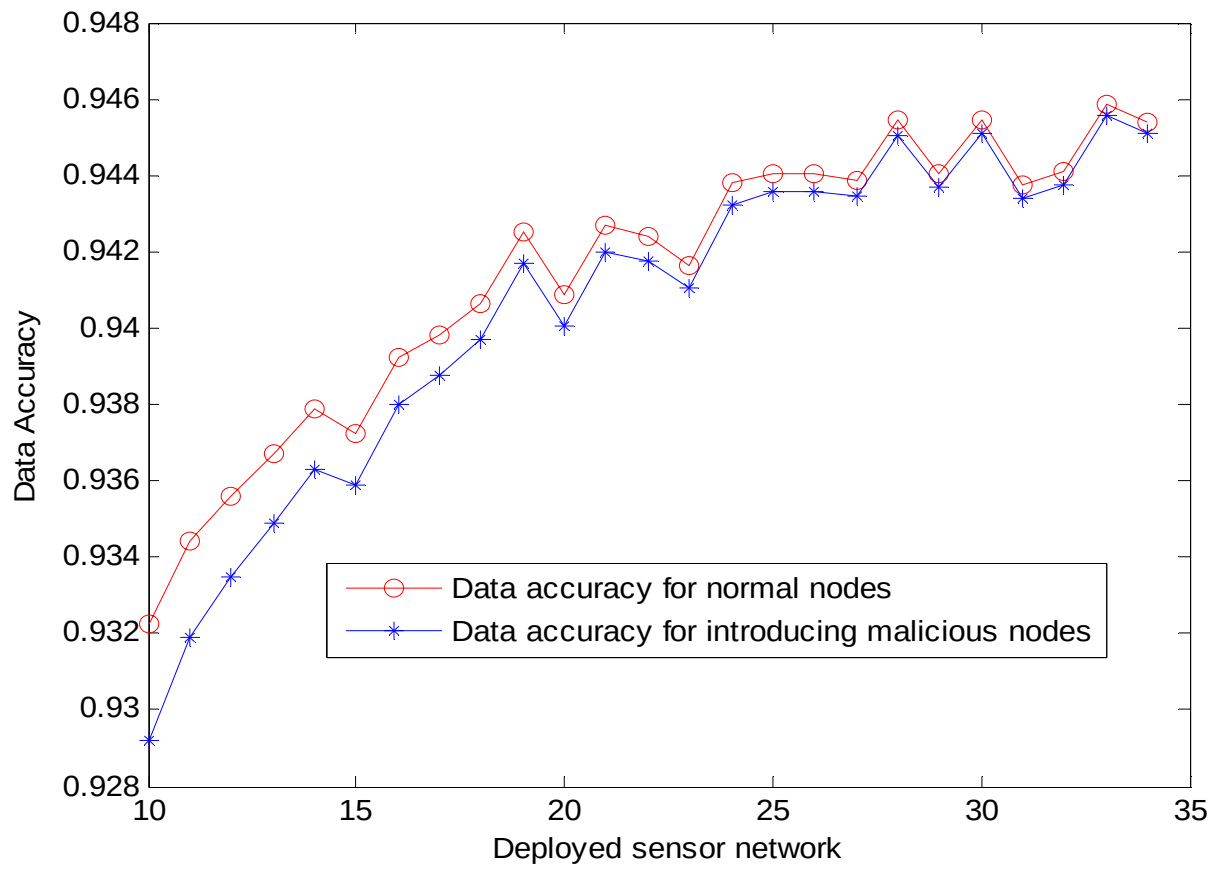

Figure 7: Comparison for data accuracy under normal nodes and malicious nodes condition in the network

## IV. PROBABILISTIC MODEL FOR SELECTING MINIMAL SET OF SENSOR NODES IN NETWORK

In the previous section, we find the estimated data accuracy for $V$ set of sensor nodes in the network and conclude that there exists an optimal $W$ set of sensor nodes which are sufficient to achieve approximately the same data accuracy. Here we demonstrate a probabilistic approach for selecting $W$ ,an optimal set of sensor nodes from $V$ set of sensor nodes, which are in active mode and keeping rest of the sensor nodes in sleep mode in the network. To find the optimal set of sensor nodes, we calculate the expectation of $V$ set of sensor nodes in the network as follows

$$E[V] = \sum_{i=1}^{v} E[\chi_i] \qquad (18)$$

Where $\chi_i \in \{0,1\}$ with $P(\chi_i = 1) = P$, for $\chi_i = 1$ denotes node $i$ is selected. We perform all the combination of $v$ sensor nodes taken $w$ at a time given as ${}^{v}C_{w}$ to perform the estimated data accuracy at the sink node. Hence the probability by which selecting an optimal number of active sensor nodes subjected to estimated data accuracy function is given in the network as

$$A(P) = \sum_{w=1}^{v} A(v = w)\, {}^{v}C_{w} P^{w} (1-P)^{v-w} \qquad (19)$$

A minimum probability ($P_{\min}$) with $\delta$ as a user dependent factor $(0 < \delta < 1)$ is calculated to find $w$ optimal number of active sensor nodes in $v$ network. Thus the minimum probability [7] for achieving a required level of data accuracy to find optimal set of sensor nodes in the network is given as

$$P_{\min} = \arg\min_{P} \{A(P) \geq \delta A_{\max}\} \quad \text{where } A_{\max} = A(1) \qquad (20)$$

In Fig. 8, we plot for the selection of optimal sensor nodes with probability $P$ with respect to estimated data accuracy function $A(P)$. Here we take $P_{\min}$ =0.2 for achieving a required level of $A(P)$ for $\delta = 0.946$ to find $W$ $(w = 7)$ optimal set of sensor nodes from $V$ $(v = 34)$ set of wake up sensor nodes in the network. Hence we get $w = 7$ optimal number of sensor nodes from $v = 34$ sensor nodes to perform the estimated data accuracy in the sink node and rest $(v - w = 27)$ of the sensor nodes goes to sleep mode. Thus we can increase the lifetime of the network by reducing the number of sensor nodes subjected to data accuracy.

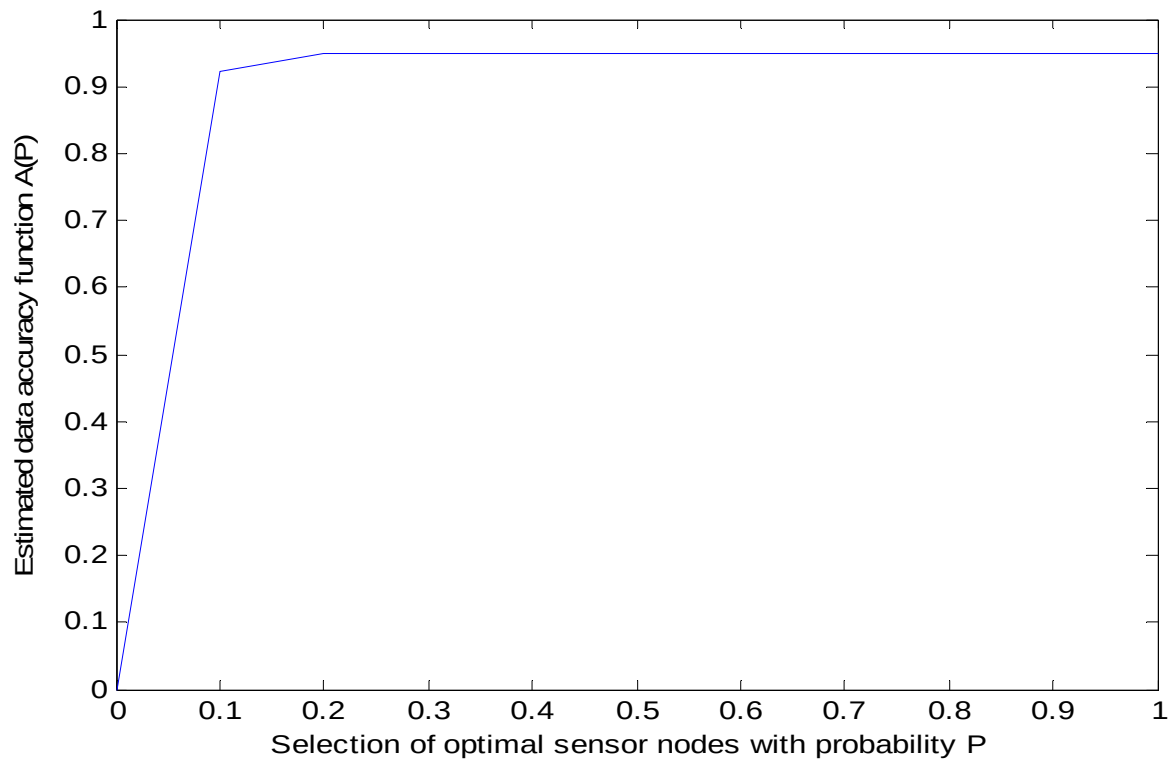

Figure 8: Selection of optimal sensor nodes with probability $P$ versus estimated data accuracy function

## V. CONCLIUSIONS

In this paper, we develop an estimated data accuracy model for the sensor nodes to sense more accurate data from the physical environment. Simulation results show this model can sense more accurate data than other information accuracy models and as the source event moves far apart from the sensor nodes the data accuracy decreases. Moreover we perform data accuracy under malicious nodes condition and conclude that if some of the sensor nodes get malicious in the network, it read and transmits inaccurate data which results poor data gathering at the sink node. Finally a probabilistic model is developed using our model to find the minimal set of sensor nodes which is sufficient to perform the same data accuracy level achieve by the maximal set of sensor nodes.